\documentclass[12pt]{article}
\usepackage{fancyhdr}

\usepackage{mathrsfs}
\usepackage[T1]{fontenc}
\usepackage{mathpazo}
\usepackage{setspace}
\usepackage{amsfonts}
\usepackage{amssymb}
\usepackage{amsmath}
\usepackage{epsfig}
\usepackage{latexsym}
\usepackage{color}
\usepackage{graphicx}
\usepackage{nicefrac}
\usepackage[latin1]{inputenc}
\usepackage{slashed}
\usepackage{multirow}
\usepackage{comment}
\usepackage{soul}
\usepackage{hyperref}
\usepackage[nosort]{cite}
\usepackage{datetime}

\usepackage[titletoc]{appendix}

\def\hybrid{\topmargin -20pt    \oddsidemargin 0pt
        \headheight 0pt \headsep 0pt
        \textwidth 6.25in       
        \textheight 9.25in       
        \marginparwidth .875in
        \parskip 5pt plus 1pt   \jot = 1.5ex}

\hybrid

\def\baselinestretch{1.2}

\catcode`\@=11

\def\marginnote#1{}
\newcount\hour
\newcount\minute
\newtoks\amorpm
\hour=\time\divide\hour by60
\minute=\time{\multiply\hour by60 \global\advance\minute by-\hour}
\edef\standardtime{{\ifnum\hour<12 \global\amorpm={am}%
        \else\global\amorpm={pm}\advance\hour by-12 \fi
        \ifnum\hour=0 \hour=12 \fi
        \number\hour:\ifnum\minute<10 0\fi\number\minute\the\amorpm}}
\edef\militarytime{\number\hour:\ifnum\minute<10 0\fi\number\minute}

\def\draftlabel#1{{\@bsphack\if@filesw {\let\thepage\relax
   \xdef\@gtempa{\write\@auxout{\string
      \newlabel{#1}{{\@currentlabel}{\thepage}}}}}\@gtempa
   \if@nobreak \ifvmode\nobreak\fi\fi\fi\@esphack}
        \gdef\@eqnlabel{#1}}
\def\@eqnlabel{}
\def\@vacuum{}
\def\draftmarginnote#1{\marginpar{\raggedright\scriptsize\tt#1}}

\def\draft{\oddsidemargin -.5truein
        \def\@oddfoot{\sl preliminary draft \hfil
        \rm\thepage\hfil\sl\today\quad\militarytime}
        \let\@evenfoot\@oddfoot \overfullrule 3pt
        \let\label=\draftlabel
        \let\marginnote=\draftmarginnote
   \def\@eqnnum{(\theequation)\rlap{\kern\marginparsep\tt\@eqnlabel}%
\global\let\@eqnlabel\@vacuum}  }

\def\preprint{\twocolumn\sloppy\flushbottom\parindent 2em
        \leftmargini 2em\leftmarginv .5em\leftmarginvi .5em
        \oddsidemargin -.5in    \evensidemargin -.5in
        \columnsep .4in \footheight 0pt
        \textwidth 10.in        \topmargin  -.4in
        \headheight 12pt \topskip .4in
        \textheight 6.9in \footskip 0pt
        \def\@oddhead{\thepage\hfil\addtocounter{page}{1}\thepage}
        \let\@evenhead\@oddhead \def\@oddfoot{} \def\@evenfoot{} }

\def\numberbysection{\@addtoreset{equation}{section}
        \def\theequation{\thesection.\arabic{equation}}}

\def\underline#1{\relax\ifmmode\@@underline#1\else
        $\@@underline{\hbox{#1}}$\relax\fi}

\def\titlepage{\@restonecolfalse\if@twocolumn\@restonecoltrue\onecolumn
     \else \newpage \fi \thispagestyle{empty}\c@page\z@
        \def\thefootnote{\fnsymbol{footnote}} }

\def\endtitlepage{\if@restonecol\twocolumn \else \newpage \fi
        \def\thefootnote{\arabic{footnote}}
        \setcounter{footnote}{0}}  

\catcode`@=12
\relax

\def\figcap{\section*{Figure Captions\markboth
        {FIGURECAPTIONS}{FIGURECAPTIONS}}\list
        {Figure \arabic{enumi}:\hfill}{\settowidth\labelwidth{Figure
999:}
        \leftmargin\labelwidth
        \advance\leftmargin\labelsep\usecounter{enumi}}}
 \relax
\def\tablecap{\section*{Table Captions\markboth
        {TABLECAPTIONS}{TABLECAPTIONS}}\list
        {Table \arabic{enumi}:\hfill}{\settowidth\labelwidth{Table
999:}
        \leftmargin\labelwidth
        \advance\leftmargin\labelsep\usecounter{enumi}}}
 \relax
\def\reflist{\section*{References\markboth
        {REFLIST}{REFLIST}}\list
        {[\arabic{enumi}]\hfill}{\settowidth\labelwidth{[999]}
        \leftmargin\labelwidth
        \advance\leftmargin\labelsep\usecounter{enumi}}}
 \relax
\makeatletter
\newcounter{pubctr}
\def\publist{\@ifnextchar[{\@publist}{\@@publist}}
\def\@publist[#1]{\list
        {[\arabic{pubctr}]\hfill}{\settowidth\labelwidth{[999]}
        \leftmargin\labelwidth
        \advance\leftmargin\labelsep
        \@nmbrlisttrue\def\@listctr{pubctr}
        \setcounter{pubctr}{#1}\addtocounter{pubctr}{-1}}}
\def\@@publist{\list
        {[\arabic{pubctr}]\hfill}{\settowidth\labelwidth{[999]}
        \leftmargin\labelwidth
        \advance\leftmargin\labelsep
        \@nmbrlisttrue\def\@listctr{pubctr}}}
 \relax
\makeatother
\newskip\humongous \humongous=0pt plus 1000pt minus 1000pt

\newif\ifdtup

\relax

\def\be{\begin{equation}}
\def\ee{\end{equation}}
\def\ba{\begin{eqnarray}}
\def\ea{\end{eqnarray}}

\def\del{\partial}

\def\a{\alpha}

\def\b{\beta}

\def\g{\gamma}

\def\d{\delta}

\def\l{\lambda}

\def\s{\sigma}

\def\no{\noindent}

\def\qq{\qquad}

\def\IR{\relax{\rm I\kern-.18em R}}

\def \ha {{1\over 2}}

\def \ov {\over}

\def\IR{\relax{\rm I\kern-.18em R}}
\def\IL{\relax{\rm I\kern-.18em L}}

\def\inv{^{\raise.15ex\hbox{${\scriptscriptstyle -}$}\kern-.05em 1}}

\begin{document}

\renewcommand{\theequation}{\thesection.\arabic{equation}}
\csname @addtoreset\endcsname{equation}{section}

\newcommand{\beq}{\begin{equation}}
\newcommand{\eeq}[1]{\label{#1}\end{equation}}
\newcommand{\ber}{\begin{equation}}
\newcommand{\eer}[1]{\label{#1}\end{equation}}
\newcommand{\eqn}[1]{(\ref{#1})}
\begin{titlepage}
\begin{center}

\hfill CERN-TH-2018-104

${}$
\vskip .2 in

{\large\bf The exact $C$-function in integrable $\lambda$-deformed theories}

\vskip 0.4in

{\bf George Georgiou,$^{1,2}$ Pantelis Panopoulos,$^{1}$ Eftychia Sagkrioti},$^{1}$\\
\vskip .1 cm
{\bf Konstantinos Sfetsos$^1$\ and\ Konstantinos Siampos}$^{3}$

\vskip 0.2in

 {\em
${}^1$Department of Nuclear and Particle Physics,\\
Faculty of Physics, National and Kapodistrian University of Athens,\\
15784 Athens, Greece
}

\vskip 0.1in

{\em
${}^2$Institute of Nuclear and Particle Physics,\\
National Center for Scientific Research Demokritos,\\
Ag. Paraskevi, GR-15310 Athens, Greece
}
\vskip 0.1in

{\em
${}^3$Theoretical Physics Department, CERN, 1211 Geneva 23, Switzerland
}
\vskip 0.1in
{\footnotesize \texttt georgiou@inp.demokritos.gr, ppanopoulos@phys.uoa.gr, esagkrioti@phys.uoa.gr,\\
ksfetsos@phys.uoa.gr, konstantinos.siampos@cern.ch}


\vskip .5in
\end{center}

\centerline{\bf Abstract}

\no
By employing CFT techniques, we show how to compute  in the context of $\lambda$-deformations of current algebras and coset CFTs the exact in the deformation parameters $C$-function
for a wide class of integrable theories that interpolate between a UV and an IR point.
We explicitly consider RG flows for integrable deformations of left-right asymmetric current algebras and coset CFTs. In all cases, the derived exact $C$-functions obey all the properties asserted by Zamolodchikov's $c$-theorem in two-dimensions.

\vskip .4in
\noindent
\end{titlepage}
\vfill
\eject

\newpage

\tableofcontents

\noindent

\def\baselinestretch{1.2}
\baselineskip 20 pt
\noindent


\section{Introduction and conclusions}

\setcounter{equation}{0}
One of the central objects in two-dimensional field theory, as well as in quantum field theory in general,  is the so-called $C$-function.
This is supposed to be a positive and monotonically decreasing function as the theory flows from the UV regime to the IR regime. Another way to state this fact is to say that the flow from the UV to the IR is irreversible. At the UV and IR fixed points the $C$-function is identified with the central charge of the corresponding conformal field theory (CFT). All these properties of the $C$-function are encoded in Zamolodchikov's $c$-theorem \cite{Zamolodchikov:1986gt}.
The physical content of the $c$-theorem is that since the stress-energy tensor couples to all degrees of freedom of a theory, the $C$-function is a measure of the effective number of degrees of freedom at a certain scale. This is in accordance with the intuition that as we lower the energy scale more and more heavy degrees of freedom decouple from the low-energy dynamics of the theory leading, thus,  to a monotonically decreasing $C$-function.

To the best of our knowledge, there is no example in which the $C$-function has been genuinely computed exactly in the couplings at all points along the RG flow as the theory interpolates between a UV and an IR fixed point.
It is the goal of the present paper to provide a calculation of the exact $C$-function with a clean CFT interpretation for a wide class of (integrable) $\sigma$-models constructed in \cite{Georgiou:2016urf,Georgiou:2017jfi,Sfetsos:2014cea,Sfetsos:2017sep}
representing exact deformations of WZW and gauged WZW CFTs for general groups.
These models generalize the construction of \cite{Sfetsos:2013wia} (for the $SU(2)$ group case see also \cite{Balog:1993es}). This will be possible since the $\beta$-functions, as well as the metrics in the space of couplings are known to all-orders in perturbation theory for the models under consideration. The latter quantities, all two- and three-point correlation functions involving currents and affine primary operators, as well as the associated operator anomalous dimensions, have been computed to all-orders in the deformation parameters $\l^i$ in a series of papers \cite{Georgiou:2015nka,Georgiou:2016iom,Georgiou:2016zyo,Georgiou:2017aei,Georgiou:2017oly,Sagkrioti:2018rwg}.
This was achieved by performing low order perturbative calculations in conjunction with certain non-perturbative symmetries in the space of couplings which these theories possess \cite{Kutasov:1989dt,Kutasov:1989aw,Itsios:2014lca,Sfetsos:2014jfa,Sfetsos:2017sep}.\footnote{ {  
The $\beta$-function
for these models, that come under the name $\lambda$-deformations, have been computed exactly in $\l$ and at large level(s)} by either resumming the perturbation series \cite{Kutasov:1989dt,Gerganov:2000mt,LeClair:2001yp}  or by the use of gravitational \cite{Itsios:2014lca,Sfetsos:2014jfa} and \cite{Georgiou:2016zyo,Georgiou:2017aei,Georgiou:2017jfi} or field theoretical methods \cite{Appadu:2015nfa,Georgiou:2017aei,Georgiou:2017oly,Sfetsos:2017sep,Sagkrioti:2018rwg}.}

 In a generic quantum field theory (QFT) with couplings $\l^i$, there exists a function $C$ of the couplings obeying the following relation \cite{Zamolodchikov:1986gt}
\be
\label{kjhj}
 {\text{d} C\ov \text{d}\ln\mu^2} =\b^i \del_i C = 24G_{ij} \b^i \b^j\,.
\ee
Here, $\displaystyle \b^i = \frac{\text{d}\l^i}{\text{d}\ln\mu^2}$ and  $G_{ij}$ is the Zamolodchikov metric in the couplings space, of the operators which drive the deformation.
The solution of this equation given the appropriate boundary conditions will specify the $C$-function of our theory. Usually this can be done only up to some finite order in perturbation theory. For our models the $\b$-functions and the metrics $G_{ij}$ have been calculated to all-orders in the perturbative expansion \cite{Kutasov:1989dt,Gerganov:2000mt,LeClair:2001yp} and  \cite{Georgiou:2017jfi,Sfetsos:2017sep,Georgiou:2015nka,Georgiou:2017aei}
enabling us to find exact expressions for the $C$-functions of the theories under consideration.
On general grounds near a fixed point of the RG flow equations the relation of the $\b$-function and the $C$-function is given by
\be
\label{gencb0}
\b^i  = {1\ov 24} G^{ij} \del_j C + \cdots\,,\quad\text{where:}\quad G^{ij}=G^{-1}_{ij}\,,
\ee
suggesting a gradient flow. This is consistent with \eqn{kjhj} and becomes
an equality for the case of a single coupling.
Provided that there is a solution to the system of equations \eqn{gencb0} as equalities, then \eqn{kjhj} is certainly solved,
leading to a monotonically decreasing behaviour from the UV to the IR as required by the $c$-theorem.

The organization of this work is as follows:
In section \ref{case1}, we will focus on the models constructed in \cite{Georgiou:2017jfi}, which can be thought of as the deformation of the sum of two WZW models at different levels $k_{1,2}$, respectively. The operators driving the models away from the UV fixed point are current bilinears with one current obeying a current algebra at level $k_1$ while the other obeys a current algebra at level $k_2$. We examine the RG flows and evaluate the $C$-function as an exact function of the deformation parameters in the large $k_{1,2}$ limit and for the case of isotropic couplings $\l_{1,2}$
(for simplification we place from now on down the indices on the $\l$'s). As discussed in detail in \cite{Georgiou:2017jfi},
as soon as the perturbation is turned on the model is driven towards another fixed point in the IR. In the case where both couplings are non-zero the IR CFT involves a coset CFT. More precisely, the flow is from the UV point where the symmetry of the theory is $G_{k_1}\times G_{k_2}$
 to the IR fixed point where the symmetry of the theory is ${G_{k_1} \times G_{k_2-k_1} \ov G_{k_2}} \times G_{k_2-k_1}$.
In the case where one of the couplings is zero the IR CFT is given by the sum of two WZW models one at level $k_1$ and the other at level $k_2 - k_1$, namely the flow is from $G_{k_1}\times G_{k_2}$ to $G_{k_1}\times G_{k_2-k_1}$. In the same section we also calculate the $C$-function for the flow  from $G_{k_1}\times G_{k_2-k_1}$ to  ${G_{k_1} \times G_{k_2-k_1} \ov G_{k_2}} \times G_{k_2-k_1}$ which is realised when one of the couplings, say $\l_2$, is set to $\l_0=\sqrt{k_1\ov k_2}$.
In section \ref{coset} we consider the flow from the UV coset theory ${G_{k_1} \times G_{k_2}\ov G_{k_1+k_2}}$
to the IR coset theory ${G_{k_2-k_1} \times G_{k_1}\ov G_{k_2}}$ \cite{Sfetsos:2017sep}.
Finally, in the appendix \ref{4-coupling}, we calculate the $C$-function for anisotropic couplings.
In this theory, which has not been shown to be integrable, each of the two coupling matrices is still diagonal but has different entries in the subgroup $H$ and in the non-symmetric Einstein subspaces. We refer to this case as the four coupling case. In all the examples considered in the present work the derived exact $C$-functions obey all properties asserted by Zamolodchikov's $c$-theorem in two-dimensions.

\section{RG flows from group spaces}
\label{case1}

In this section, we will calculate the exact $C$-function for the class of theories constructed in \cite{Georgiou:2017jfi}.
We consider the following action with two coupling matrices $(\l_{1,2})_{AB}$
\begin{equation}
S_{\l_1,\l_2}=S_{k_1}(g_1)+S_{k_2}(g_2)+\frac{\sqrt{k_1k_2}}{\pi}\int \text{d}^2\s\big((\l_1)_{AB}J^A_{1+}J^B_{2-}+(\l_2)_{AB}J^A_{2-}J^B_{1+}\big)\  ,
\label{action}
\end{equation}
where the capital indices run over a semi-simple group $G$.  The effective action for this theory has been
constructed in  \cite{Georgiou:2017jfi}, but its  explicit form will not be needed for our purposes. Moreover, we will focus on
the case of isotropic couplings in which the matrices will be $(\l_{i})_{AB}=\l_i \d_{AB}$, so that we have just two parameters.
A generalization to a case with four couplings will be considered in the  appendix \ref{4-coupling}.

\no
The Zamolodchikov metric in the coupling space can be calculated to all-orders in $\l_{1,2}$ and is given in
the large $k_{1,2}$ limit by \cite{Kutasov:1989dt,Georgiou:2015nka}
\be
\label{Gij}
G_{ij}={\d_{ij}\ov 2}{\dim G\ov (1-\l_i^2)^2}\ ,
\ee
where we note that in the large-$k$ expansion, the leading term in the metric presented above depends only on the Abelian part of the current OPEs.\footnote{{ The metric takes the form \eqn{Gij} since the current bilinears $J^A_{1+}J^B_{2-}$ and $J^A_{2+}J^B_{1-}$ in \eqref{action} do not interact.  Therefore, it  is a double copy of the single-coupling case \cite{Kutasov:1989dt,Georgiou:2015nka}.}}
If we know the $\b$-function near a CFT fixed point, either at the UV or at the IR, by integrating \eqn{gencb0}
we may know near the same point the behaviour of the $C$-function, as well.  The integration constant is fixed by requiring that the $C$-function at the CFT point coincides with the central charge of the corresponding CFT.
In what follows, the exact expression for $C$ is found by symmetry arguments in conjunction with the perturbative results.

Before proceeding, it is convenient to define the combinations of the levels $k_{1,2}$ (assuming with no loss of generality that $k_2\geqslant k_1$) given by
\begin{equation*}
\l_0=\sqrt{k_1\ov k_2}\leqslant 1\ ,\qquad k= \sqrt{k_1k_2}\ .
 \end{equation*}
The running of the couplings has been computed exactly in the $\l_i$'s and to leading order for $k\gg 1$
{, by CFT  and gravitational methods in \cite{LeClair:2001yp,Georgiou:2016zyo}
and \cite{Georgiou:2017jfi}, respectively}
\be\label{betas}
\b^i(\l_i;\l_0) = -{c_G\ov 2 k} {\l_i^2 (\l_i-\l_0)(\l_i-\l_0^{-1})\ov (1-\l_i^2)^2}\,,\quad i=1,2\ ,
\ee
where $c_G$ is the quadratic Casimir for $G$ in the adjoint representation, see \eqn{fsidentity}.
This demonstrates that the flows for $\l_1$ and $\l_2$ are decoupled from each other and that there are two fixed points at the UV and at the IR at $\l_i=0$ and $\l_i=\l_0$, respectively.
Near the UV fixed point we obtain perturbatively that
\begin{equation*}
\b^i(\l_i;\l_0) =
-{c_G\ov 2 k} \l_i^2 +  {\cal O}(\l_i^3)  \ ,\quad i=1,2\ .
\end{equation*}
The central charge at the UV CFT is given by the sum of the central charges of two WZW models at levels $k_{1,2}$, respectively.
Keeping the two leading terms in the ${1/k}$ -expansion we obtain that
\be
\label{cuv}
c_{\rm UV}= {2 k_1\dim G\ov 2 k_1 + c_G}+ {2 k_2\dim G\ov 2 k_2 + c_G} = 2 \dim G
- {c_G \dim G\ov 2 k} (\l_0+\l_0^{-1}) + \cdots \ .
\ee
Therefore, using also that at leading order near $\l_i=0$, the metric
$\displaystyle G_{ij}=\ha \d_{ij} \dim G$, we have by solving \eqref{gencb0} that
\be\label{pert-C1}
C(\l_1,\l_2;\l_0) = 2 \dim G - {c_G \dim G\ov k}\left(\ha (\l_0+\l_0^{-1}) + 2 \l_1^3 + 2\l_2^3\right)+  {\cal O}(\l^4)\ .
\ee
The { effective action of the} $\s$-model { \eqref{action}, was constructed in \cite{Georgiou:2017jfi},} and in particular the $\beta$-function is { explicitly} invariant under the symmetry\footnote{
{At the level of the effective action of \eqref{action}, the symmetry \eqref{duall} is accompanied by an inversion of the group elements $g_{1,2}\to g_{2,1}^{-1}$, see Eqs.(2.12) \& (2.14) in \cite{Georgiou:2017jfi}.
This generalizes a similar symmetry found in \cite{Itsios:2014lca} for the single $\l$-deformed model \cite{Sfetsos:2013wia}.
}}
\be
\l_i\to {1\ov \l_i}\ \quad i=1,2\ ,\quad k_1\to -k_2\  ,\quad k_2\to -k_1\  .
\label{duall}
\ee
Note that, under this transformation the expansion parameter $k\to -k$.
The two fixed points of the above symmetry for the parameters $\l_i$ are at $\l_i=\pm 1$.
In order to have  well behaved expansions of \eqref{betas} around these fixed points we expand
\begin{equation}
\label{nonabps}
\l_i= \pm 1 -{b_i\ov k^{1/3}}\ ,\quad k\to \infty\ ,\quad \l_0={\rm fixed}\, .
\end{equation}
Hence, as explained in \cite{Georgiou:2015nka}, the $C$-function to ${\cal O}(1/k)$ should be of the following form
\be\label{C1}
C(\l_1,\l_2;\l_0) = 2\dim G - {c_G \dim G\ov 4 k}\left({f(\l_1;\l_0)\ov (1-\l_1^2)^3}
+{f(\l_2;\l_0)\ov (1-\l_2^2)^3}\right)\ ,
\ee
with the analytic function $f(\l;\l_0)$ obeying, thanks to \eqn{duall},
the condition $f(\l;\l_0)= \l^6 f(\l^{-1};\l_0^{-1})$.
Hence, $f(\l;\l_0)$ should be a six-degree polynomial in $\l$ with coefficients that will generically
depend on $\l_0$. Matching with the perturbative results \eqref{pert-C1} gives
\be
\label{cfall}
f(\l;\l_0)= (\l_0+\l_0^{-1})(1-3 \l^2 - 3 \l^4 + \l^6) + 8 \l^3\ .
\ee
Note that $C(\l_1^{-1},\l_2^{-1};\l_0^{-1})=C(\l_1,\l_2;\l_0)$, implying
invariance under \eqn{duall}.\footnote{{
A method of deriving what has been called effective central charge is by employing TBA techniques for determining the ground state energy of the system, initially put forward in \cite{Zamolodchikov:1991vh,Zamolodchikov:2000kt}.}}

\subsection*{A shortcut}

Using \eqn{kjhj} we obtain that
\begin{equation}
\label{kjhj0}
 \del_{\l} C = 24 G_{\l\l} \b^\l\ .
\end{equation}
This relation is exact in $\l$ and should hold for each of the two couplings $\l_{1,2}$, separately. Then, for our case
\begin{equation}
\label{shortcc}
\begin{split}
& C =  c_{\rm UV}  -6 {c_G \dim G\ov k} \int_0^{\l_1} \text{d}x {x^2 (x-\l_0)(x-\l_0^{-1})\ov (1-x^2)^4 }-
(\l_1\to\l_2)
\\
&\phantom{x}
= c_{\rm UV} -{c_G \dim G\ov 2k}\l_1^3 {4 -\l_1(3-\l_1^2) (\l_0+\l_0^{-1})\ov (1-\l_1^2)^3}-(\l_1\to\l_2)\ ,
\end{split}
\end{equation}
where $c_{\rm UV}$ is given by \eqn{cuv}. After some algebra we obtain indeed \eqn{C1} and
\eqn{cfall}. Note that the term arising from the integration is not by itself invariant under \eqn{duall}.

\subsection*{The equal level case}

In the equal level case, i.e. when $\l_0=1$, the $C$-function \eqref{C1} becomes (we also take
$\l_1=\l_2=\l$)
\begin{equation}
 \label{Kutasov}
C(\l) = 2\dim G - {c_G \dim G\ov k} {1+2 \l  + 2 \l^3 + \l^4\ov  (1-\l)(1+\l)^3}\ .
\end{equation}
This is the $C$-function corresponding to the simplest $\l$-deformed model of \cite{Sfetsos:2013wia}.
This  becomes strongly coupled at $\l=\pm 1$. As shown in \cite{Sfetsos:2013wia} and in
\cite{Georgiou:2016iom} what makes sense near these points are the non-Abelian T-duality and pseudodual model limits
\eqn{nonabps}, for $\l=1$ and $\l=-1$,  respectively. In these limits $C(\l)$ remains indeed finite.

\no
Let us mention that the $C$-function for this particular case, known also as  the non-Abelian bosonized Thirring model, was implicitly derived in \cite{Kutasov:1989dt,Kutasov:1989aw}, albeit not in the invariant form \eqn{Kutasov}.
Indeed, the effective potential calculated in
\cite{Kutasov:1989dt} satisfies the same differential equation as the $C$-function. However, unlike this, the effective potential is not invariant under the non-perturbative symmetry $\l \rightarrow 1/\l,\,\,k\rightarrow -k$.\footnote{
For certain integrable deformations of the isotropic principal chiral model
\cite{Delduc:2014uaa},
the Weyl anomaly coefficient for $\sigma$-model in the NS--NS sector   \cite{ Tseytlin:1987bz,Tseytlin:2006ak} was worked out \cite{Demulder:2017zhz}. This can be interpreted as a "generalized central function". However, as stressed in  \cite{Tseytlin:1987bz,Tseytlin:2006ak} it can not be in general identified with the $C$-function of Zamolodchikov since its flow is not generically monotonic due to the indefinite metric in the space of couplings $(g_{\mu\nu},B_{\mu\nu},\Phi)$. It seems
worth pursuing investigations in this direction.}

\subsection{From $G_{k_1} \times G_{k_2}$  to ${G_{k_1} \times G_{k_2-k_1} \ov G_{k_2}} \times G_{k_2-k_1}$}

Expanding \eqref{C1} around the IR fixed point at $\l_1=\l_2=\l_0$ we obtain that
\be
\begin{split}
& C(\l_1,\l_2;\l_0)= 2\dim G- {c_G \dim G\ov  k} {1+\l_0^4\ov 2\l_0 (1-\l_0^2)}+{\cal O}(\l-\l_0)^2\,.
\end{split}
\label{dh12}
\ee
The leading term should correspond to the large $k$ expansion of the central charge of the IR CFT which was explicitly identified in \cite{Georgiou:2017jfi} to be the coset ${G_{k_1} \times G_{k_2-k_1} \ov G_{k_2}} \times G_{k_2-k_1}$.\footnote{It can be easily seen that the central charge of ${G_{k_1} \times G_{k_2-k_1} \ov G_{k_2}} \times G_{k_2-k_1}$ is
invariant under the extension of the  symmetry \eqn{duall} which involves the shifts of levels
\begin{equation*}
k_1\to -k_2- c_G\  ,\quad k_2\to -k_1- c_G\  ,
\end{equation*}
valid beyond the large $k$ limit. The transformation of $\l$ is probably not simply an inversion as
in \eqn{duall}, but it may involve $k$-corrections as well.}
Its central charge for $k\gg 1$ gives indeed  $C(\l_0,\l_0;\l_0)$, i.e. the leading term in \eqn{dh12}, as it should be.
This consists a non-trivial check of our exact formula \eqref{C1}.

\subsection{From $G_{k_1} \times G_{k_2}$ to $G_{k_1} \times G_{k_2-k_1}  $}

One may consistently set one of the couplings to zero, say $\l_2=0$, as it can be seen from the expressions of
the $\b$-functions of the theory \eqref{betas}.
Renaming $\l_1$ to $\l$ and attending this to the general expression of \eqref{C1}, \eqref{cfall} we obtain that
\be
\label{jh33}
C(\l;\l_0)=  2\dim G - {c_G \dim G\ov 2 k}{(\l_0+\l_0^{-1})(1-3 \l^2) +4  \l^3 \ov (1-\l^2)^3}\ .
\ee
Expanding around the IR fixed point at $\l=\l_0$ one has that
\be
\begin{split}
& C(\l;\l_0)= 2\dim G- {c_G \dim G\ov  2 k} {1\ov  \l_0(1-\l_0^2)}+{\cal O}(\l-\l_0)^2\,.
\end{split}
\label{dh122}
\ee
The leading term should correspond to the large $k$ expansion of the central charge of the IR CFT, which consists of
a sum of two WZW models at levels $k_1$ and $k_2-k_1$ \cite{Georgiou:2017jfi}.
For $k\gg 1$ we get indeed $C(\l_0;\l_0)$, i.e. the leading term in \eqn{dh122}, as it should be.
Note also that
\eqn{jh33} is not invariant under $\l\to 1/\l$ and $k\to -k$.
This is not a surprise in the sense that at the end point of the flow the
central charge $c_{\rm IR}$ is not invariant under this symmetry.

\subsection{From $G_{k_1}\times G_{k_2-k_1}$ to  ${G_{k_1} \times G_{k_2-k_1} \ov G_{k_2}} \times G_{k_2-k_1}$}

There is another consistent truncation in the coupling space. This is to set one of the couplings, say $\l_2$ to $\l_0$
 and rename $\l_1$
as $\l$. Plugging these to the general expression for $C(\l_1,\l_2;\l_0)$ \eqref{C1} we obtain that
\be
C(\l;\l_0)=  2\dim G - {c_G \dim G\ov 2 k}
{1-3 \l^2 +\l_0^4 \l^4(3 -\l^2) +4\l_0(1-\l_0^2)\l^3
  \ov\l_0 (1-\l_0^2) (1-\l^2)^3}\ .
\ee
Near the UV fixed point at $\l=0$ we obtain the leading term in \eqref{dh122}, as expected since the IR fixed point of the flow described in the previous section is the UV fixed point of the present flow.
Similarly, near the IR fixed point at $\l=\l_0$ we retrieve \eqref{dh12}, as one should expect.
This $C$-function is invariant under $\l\to 1/\l$ and $k\to -k$.

\section{RG flows from coset spaces}
\label{coset}

In this section, we evaluate the exact $C$-function for the flow from the coset theory with symmetry ${G_{k_1} \times G_{k_2}\ov G_{k_1+k_2}}$ (UV fixed point) to the coset theory with symmetry ${G_{k_2-k_1} \times G_{k_1}\ov G_{k_2}}$ (IR fixed point).\footnote{Other related works studying aspects of RG flows of coset theories include \cite{Bernard:1990cw,Crnkovic:1989gy,Ahn1990,Zam-gen,Ravani}.} These theories were constructed first time in \cite{Sfetsos:2014cea} for $G=SU(2)$ where an explicit action was given.
The effective action for this, taken into account all loop
effects in the $\l_i$'s integrability properties, as well as the running of the coupling were subsequently examined for a
general group $G$ in \cite{Sfetsos:2017sep}, where the corresponding RG flow was also identified.

We start by defining the parameters
\begin{equation*}
s_i= {k_i\ov k_1+k_2}\ ,\quad i=1,2\
\end{equation*}
and
\begin{equation*}
\l_1^{-1}={s_2-3s_1}\,,\quad \l_2^{-1}={s_1-3s_2}\,,\quad
 \l_3^{-1}=(s_1-s_2)^2=1-4 s_1 s_2\ , \quad \l_f^{-1}=  1-8 s_1s_2\ .
\end{equation*}
The one-loop $\beta$-function exact in $\l$ and to leading order for $k\gg 1$ is given \cite{Sfetsos:2017sep}
by \footnote{\label{symspas}
For equal levels
$k_2=k_2=k$ we have that $\l_1=\l_2=\l_f=-1$ and $\l_3\to \infty$. Then,
$
\b_\l = -{c_G\ov 4k }\l,
$
that is the perturbative example is also exact (for $k\gg 1$) as found in \cite{Itsios:2014lca,Appadu:2015nfa} for all
symmetric spaces. }
\begin{equation}
\beta_\l=-\frac{c_G\l(1-\l_1^{-1}\l)(1-\l_2^{-1}\l)(1-\l^{-1}_3\l)} {2(k_1+k_2)
(1-\l_f^{-1}\l)^2}\,.
\end{equation}
The central charge at the UV CFT ${G_{k_1} \times G_{k_2}\ov G_{k_1+k_2}}$ is easily obtained to be
\be
 c_{\rm UV}=   \dim G - {c_G \dim G\ov  k} {1+\l_0^2+\l_0^4\ov 2\l_0(1+\l_0^2)} + \cdots \ .
\ee
The central charge at the IR CFT ${G_{k_2-k_1} \times G_{k_1}\ov G_{k_2}}$ is when $\l=\l_2<0$ and is given by
\be
\label{cirr}
c_{\rm IR}= \dim G - {c_G \dim G\ov  k} {1-\l_0^2+\l_0^4\ov 2\l_0(1-\l_0^2)} + \cdots \ .
\ee

\no
The action of the theory is invariant  under the following remarkable symmetry transformation \cite{Sfetsos:2017sep}
\begin{equation}
\label{symmetry}
\l\to
 \frac{1-(s_1-s_2)^2\l}{(s_1-s_2)^2-(1-8s_1s_2)\l}\ ,\quad k_i\to -k_i\ ,\quad i=1,2\
\end{equation}
and the same is true for the $\b$-function.
This symmetry has
two fixed points for the parameter $\l$ given by
$\l=1$ and $\l=\l_f$. Near these points the limits
\begin{equation*}
\l =1 -{b\ov k} \,,\quad k\to \infty\,,
\end{equation*}
and
\begin{equation*}
\l= \l_f -\frac{b}{k}\ ,\quad \l_0=1-\frac{n}{2 k}\,, \quad k\to \infty\,,
\end{equation*}
of all quantities should be well defined (as they are for the $\b$-function).

\no
The metric in the space of couplings is given by
\be
\label{dfh11}
G_{\l\l}={8 s_1^2 s_2^2 \dim G \ov (1-\l)^2(1-\l^{-1}_f\l)^2}\ .
\ee
It has the correct behaviour being singular at the fixed points of the symmetry
transformation and also for $k_1=k_2$ we get $\l_f=-1$ reducing, thus, to the known one, namely $G_{\l\l}={ \dim G \ov 2 (1-\l^2)^2}$.
Furthermore, the symmetry \eqn{symmetry} leaves invariant the line element with metric \eqn{dfh11}.
Finally, the overall coefficient is chosen so that the correct central charge in the IR is reproduced.
It will be interesting to derive \eqn{dfh11} along the lines of \cite{Kutasov:1989dt,Georgiou:2015nka}.

\no
To obtain { the exact in $\l$ and to ${\cal O}(1/k)$ one solves \eqn{kjhj0}}
\ba
&& C(\l;\l_0) =
 \dim G - {c_G \dim G\ov  k} {1+\l_0^2+\l_0^4\ov 2\l_0(1+\l_0^2)}
\\
&&\phantom{x}
  -{16c_G \dim G\ov k} {\l_0^5\l^2\ov (1+\l_0^2)^7}
  {3(1+\l_0^2)^2 +2 (1+10 \l_0^2+\l_0^4)\l -(5-22\l_0^2+5\l_0^4)\l^2\ov
  (1-\l)(1-\l_f^{-1}\l)^3}\ .
  \nonumber
\ea
At the IR fixed point situated at $\l=\l_2$ we find \eqn{cirr}, as it should be.
In addition, $ C(\l;\l_0)$ is invariant under the symmetry \eqn{symmetry}.\footnote{
We note that in the equal level case, cf. footnote \ref{symspas},  the $C$-function drastically simplifies to
\begin{equation*}
C(\l) = \dim G - {3 c_G \dim G\ov 4 k} {1+\l^2\ov 1-\l^2}\ .
\end{equation*}

}

\appendix

\section{The case of four couplings}\label{4-coupling}

\subsection*{Preliminaries}

In this appendix, we consider again the class of models presented in \cite{Georgiou:2017jfi}, but with the couplings matrices being more general. Namely, let's consider again the action \eqn{action}
and split the semi-simple group $G$ into one of its subgroups $H$ and the corresponding coset space $G/H$.
In the split index $A=(a,\a)$, Latin (Greek) letters denote subgroup (coset) indices. Then for
non-symmetric Einstein-spaces $G/H$ \cite{Forgacs:1985vp,Lust:1986ix} we have that
\be
\label{fsidentity}
\begin{split}
&f_{ACD} f_{BCD} = c_G \d_{AB}\ ,\quad f_{acd} f_{bcd} = c_H \d_{ab}\ ,
\quad
f_{\a\g\d}f_{\b\g\d}=c_{G/H} \d_{\a\b}\ ,
\\
&f_{a\g \d} f_{b\g \d} = (c_{G}-c_{H})\d_{ab}\,,\quad f_{\a\g c} f_{\b\g c} = \ha (c_{G} -c_{G/H})  \d_{\a\b}\ .
\end{split}
\ee
Among the above identities, the non-trivial one that essentially defines a non-symmetric Einstein space, is the one involving
purely Greek indices.

\no
We will work for diagonal deformation matrices  $(\l_i)_{AB}=\l_{H_i}\,\d_{ab}+\l_i\,\d_{\a\b}$, where $\l_{H_i}$ and $\l_i$ denote the subgroup and coset deformation parameters respectively.
The $\b$-functions for each $(\l_i)_{AB}$ are given by \cite{Sagkrioti:2018rwg}
\begin{equation*}
\begin{split}
&
\b_{\l_H} = -{(\l_H-\l_0)(\l_H-\l_0^{-1})\ov 2 k}
\left( c_H {\l^2_H \ov (1-\l_H^2)^2} +
(c_G-c_H) {\l^2 \ov (1-\l^2)^2} \right)\ ,
\\
&
\b_\l =
 -{1\ov 2 k} \left(c_{G/H} {\l^2 (\l-\l_0)(\l-\l_0^{-1})\ov (1-\l^2)^2}
 + {c_G-c_{G/H}\ov 2}\right.
 \\
 & \left. \times {\l \ov (1-\l^2) (1-\l_H^2)}
\left((\l_0^{-1}-\l_H)(\l_0\l_H-\l^2)+ (\l_0-\l_H)(\l_0^{-1}\l_H-\l^2)\right) \right)\ .
\end{split}
\end{equation*}
These $\b$-functions  are invariant under
\begin{equation*}
\l\to\l^{-1}\ , \quad \l_H\to\l_H^{-1} \ ,\quad  k_{1,2}\to-k_{2,1} \ ,
\end{equation*}
and have as fixed points
\be
(\l_H,\l)=\{(0,0)\,,(\l_0,\l_0)\,,(\l_0,0)\}\,.
\ee
In the case at hand, the Zamolodchikov metric in the space of couplings $G_{ij }$ reads
\begin{equation}
G_{ij}=\frac{1}{2}\bigg(\begin{matrix}
G^{(1)}_{ij} && 0\\
0 && G^{(2)}_{ij}
\end{matrix} \bigg)\ ,\quad G^{(i)}_{ij}=\begin{pmatrix}
\frac{\text{dimG/H}}{(1-\l_i^2)^2}
&& 0\\
0 && \frac{\text{dimH}}{(1-\l_{H_i}^2)^2}
\end{pmatrix}\ ,\quad i=1,2\ .
 \label{metric4couplings}
\end{equation}

\subsection*{The $C$-function}

We will compute the general $C$-function corresponding to the action \eqref{action}. From \eqref{gencb0} with the use of \eqref{metric4couplings} we end up with 4 differential equations. However, from the form of the two $\b$-functions we know that the couplings $(\l_1)_{AB}$ and $(\l_2)_{AB}$ are independent to each other. Hence, the occurring differential equations are pairwise decoupled.

The general solution of these equations is given by
\begin{equation}
C(\l_1,\l_{H_1},\l_2,\l_{H_2};\l_0)=2 \dim G -
 \frac{1}{4 k}\Big(\tilde{C}(\l_1,\l_{H_1},\l_0)+\tilde{C}(\l_2,\l_{H_2};\l_0)\Big)\,, \label{c-function_4couplings}
\end{equation}
where
\begin{eqnarray*}
\tilde{C}(\l,\l_{H};\l_0)=\frac{1}{(1-\l^2)^3}   \bigg(  c_G\,\text{dimG}\,g_1+c_G\ \text{dimH}\,g_2 +c_H\,\text{dimH}\,g_3 \bigg)\,,
\end{eqnarray*}
where the functions $g_i(\l,\l_{H};\l_0)$, $i=1,2,3$ are given by
\begin{align*}
\begin{split}
&g_1=(\l_0+\l_0^{-1})(1-3\l^2-3\l^4+\l^6)+8\l^3\ ,
\\
&g_2=12\l^2(\l_0+\l_0^{-1}-2\l)-\frac{12\l^2(1-\l^2)}{(1-\l_{H}^2)}(\l_0+\l_0^{-1}-2\l_{H}) \ ,
\\
&g_3=-2\big((\l_0+\l_0^{-1})(1+3\l^2)-8\l^3\big)+\frac{2(1-\l^2)}{(1-\l_{H}^2)^3}\bigg[(\l_0+\l_0^{-1})\big(1+4\l^2
\\
& \qq+\l^4 -3(1+\l^2)^2\l_H^2+6\l^2\l_H^4\big)+4(1+4\l^2+\l^4)\l_H^3-12\l^2\l_H(1+\l_H^4)\bigg]\ .
\end{split}
\end{align*}
In order to find \eqref{c-function_4couplings}, we used the consistency relation
\begin{equation*}
c_{G/H}=c_G-\frac{2\text{dimH}(c_G-c_H)}{\text{dimG/H}}\ ,
\end{equation*}
which can be easily proved by appropriately tracing the free indices of the second line in \eqref{fsidentity}.
The function \eqref{c-function_4couplings} indeed gives the correct UV and IR values for the central charges \eqref{cuv} and \eqref{dh12}. It remains invariant under the symmetry $(\l_i\to\l_i^{-1},\l_{H_i}\to\l_{H_i}^{-1},\l_0\to\l_0^{-1},k\to -k)$.
The $C$-function we have derived, satisfies \eqref{kjhj}, has the right behaviour when $\l=\l_H$ or $\l=0$ and monotonically decreases as one transverses from the UV to the IR.

\subsubsection*{Other important RG flows}

\begin{enumerate}

\item

If we take $\l_2=\l_{H_2}=0$ and $\l_1=\l_{H_1}=\l$, \eqref{c-function_4couplings} takes the form
\begin{equation*}
C(\l,\l,0,0;\l_0)=2\text{dimG}-\frac{c_G\text{dimG}}{2k}\frac{(\l_0+\l_0^{-1})(1-3\l^2)+4\l^3}{(1-\l^2)^3}
\end{equation*}
which is invariant up to a non-vanishing constant under the duality-type symmetry and has the correct IR central charge at $\l=\l_0$,
corresponding to the flow from $G_{k_1}\times G_{k_2}$ to $G_{k_1}\times G_{k_2-k_1}$.

\item
By taking $\l_2=\l_{H_2}=\l_0$ and $\l_1=\l_{H_1}=\l$, \eqref{c-function_4couplings} reduces to
\begin{equation*}
C(\l,\l,\l_0,\l_0;\l_0)=2\text{dimG}-\frac{c_G\text{dimG}}{2k}\frac{1-3\l^2+4\l_0^4\l^4(3-\l^2)+\l_0(1-\l_0^2)\l^3}{\l_0(1-\l_0^2)(1-\l^2)^3}
\end{equation*}
representing the flow from $G_{k_1}\times G_{k_2-k_1}$ to $\frac{G_{k_1}\times G_{k_2-k_1}}{G_{k_2}}\times G_{k_2-k_1}$, with the correct IR central charge at $\l=\l_0$.

\item
Taking $\l_{1,2}=0$ and $\l_{H_1}=\l_{H_2}=\l$, \eqref{c-function_4couplings} reduces to
\begin{equation*}
C(0,\l,0,\l;\l_0)=2\text{dimG}-\frac{c_G \text{dimG}}{2k}\left(\l_0+\l_0^{-1}\right)
-\frac{c_H\text{dimH}}{k}\frac{\l^3(4+\l(\l_0+\l_0^{-1})(\l^2-3)}{(1-\l^2)^3}\ .
\end{equation*}

\item
Finally, considering $\l_{1,2}=0=\l_{H_2}$ and $\l_{H_1}=\l$, \eqref{c-function_4couplings} reduces to
\begin{equation*}
C(0,\l,0,\l;\l_0)=2\text{dimG}-\frac{c_G \text{dimG}}{2k}\left(\l_0+\l_0^{-1}\right)
-\frac{c_H\text{dimH}}{2k}\frac{\l^3(4+\l(\l_0+\l_0^{-1})(\l^2-3)}{(1-\l^2)^3}\ .
\end{equation*}

\end{enumerate}
It will be very interesting to identify the IR CFTs for the cases 3 and 4 above.

 \end{document}